\documentclass[11pt,oneside]{extarticle}

\usepackage[a4paper,margin=3cm]{geometry}

\usepackage{xcolor}
\definecolor{celestezblue}{HTML}{2C7DF7}

\usepackage{fontspec}
\defaultfontfeatures{Ligatures=TeX}

\usepackage{unicode-math}
\usepackage{microtype}
\usepackage{setspace}
\usepackage{csquotes}
\usepackage{booktabs}
\usepackage{amsmath}
\usepackage{enumitem}
\usepackage{graphicx}
\usepackage{hyperref}
\usepackage{datetime2}
\usepackage{titlesec}
\usepackage{fancyhdr}
\usepackage{tikz}
\usetikzlibrary{positioning,arrows.meta,calc,fit}
\usepackage{listings}
\lstdefinestyle{AppendixCode}{basicstyle=\ttfamily\footnotesize,frame=single,frameround=tttt,rulecolor=\color{celestezblue},backgroundcolor=\color{black!2},keywordstyle=\color{celestezblue}\bfseries,commentstyle=\color{gray},stringstyle=\color{green!40!black},showstringspaces=false,breaklines=true,columns=fullflexible}
\newcommand{\apicode}[1]{{\ttfamily\color{celestezblue}#1}}
\usepackage{tabularx}
\usepackage{pgfplots}
\pgfplotsset{compat=1.18}
\newcolumntype{Y}{>{\raggedright\arraybackslash}X}
\usepackage[numbers,sort&compress]{natbib}

\newcommand{\LogoTargetHeight}{1.45cm}

\newcommand{\LogoGraphic}{\includegraphics[height=\LogoTargetHeight,width=\linewidth,keepaspectratio]{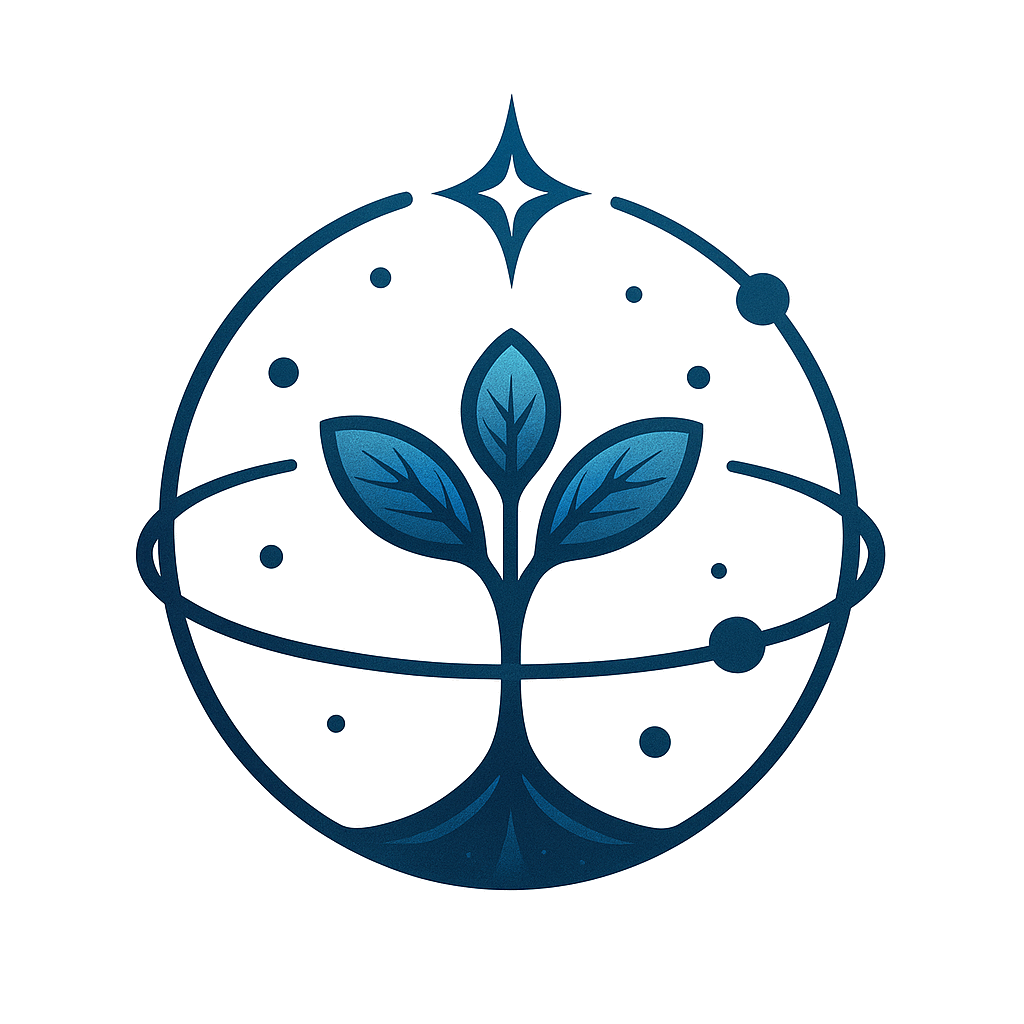}}

\newcommand{\ProjectName}{Canonical LST{}}
\newcommand{\TokenTicker}{sTEZ{}}
\newcommand{\ExchangeRateSymbol}{\ensuremath{\mathit{R}_t}}
\newcommand{\LedgerBalance}{\ensuremath{\mathit{L}_t}}
\newcommand{\TokenSupply}{\ensuremath{\mathit{S}_t}}
\newcommand{\FXRate}{\ensuremath{\mathit{FX}_t}} 
\newcommand{\ExchangeRate}{\ensuremath{\ExchangeRateSymbol = \tfrac{\LedgerBalance}{\TokenSupply}}} 
\usepackage{mdframed}
\usepackage{capt-of} 

\hypersetup{
  colorlinks=true,
  linkcolor=celestezblue,
  citecolor=celestezblue,
  urlcolor=celestezblue,
  pdftitle={\ProjectName Whitepaper},
  pdfauthor={\ProjectName Team},
  pdfsubject={Protocol Design},
  pdfkeywords={blockchain, tezos, protocol, consensus}
}

\titleformat{\section}{\Large\bfseries\sffamily}{\thesection}{1em}{}
\titleformat{\subsection}{\large\sffamily\bfseries}{\thesubsection}{1em}{}
\setlist{itemsep=0.25em,topsep=0.25em}

\usepackage{needspace}
\usepackage{float}
\usepackage[automake,style=long,nonumberlist]{glossaries}
\newglossary[protidx]{protocol}{protacr}{protgls}{Protocol and Network Concepts}
\newglossary[mechidx]{mechanics}{mechacr}{mechgls}{\ProjectName\ and Staking Mechanics}
\newglossary[finidx]{finance}{finacr}{fingls}{Financial and Valuation Terms}
\newglossary[regidx]{regulatory}{regacr}{reggls}{Regulatory and Compliance}
\newglossary[techidx]{tech}{techacr}{techgls}{Technical / Operational}

\newglossaryentry{pos}{type=protocol,name=Proof-of-Stake (PoS),description={Consensus model where validators stake tez to propose/attest blocks and can be slashed for misbehaviour}}
\newglossaryentry{consensus}{type=protocol,name=Consensus,description={Protocol process by which the network agrees on the next valid block}}
\newglossaryentry{block}{type=protocol,name=Block,description={Ordered container of operations appended to the chain}}
\newglossaryentry{blockchain}{type=protocol,name=Blockchain,description={Append-only, cryptographically linked sequence of blocks forming the shared ledger}}
\newglossaryentry{cycle}{type=protocol,name=Cycle,description={Epoch boundary used for reward accrual, allocation updates, and governance timing}}
\newglossaryentry{validator}{type=protocol,name=Validator,description={Block producer / attester (historically a baker) participating in PoS consensus}}
\newglossaryentry{tezosprotocol}{type=protocol,name=Tezos protocol,description={Layer-1 protocol defining consensus, governance (amendments), execution, and economic rules}}
\newglossaryentry{attestation}{type=protocol,name=Attestation,description={Validator vote signalling support for a proposed block; missed or conflicting attestations impact eligibility}}
\newglossaryentry{denunciation}{type=protocol,name=Denunciation,description={On-chain report of a slashable offense (double-baking or double-attesting) triggering penalties}}
\newglossaryentry{doublebaking}{type=protocol,name=Double-baking,description={Producing two different blocks at the same level; a slashable consensus fault}}
\newglossaryentry{amendment}{type=protocol,name=Amendment,description={Formal governance process upgrading protocol code and parameters via multi-phase voting}}
\newglossaryentry{activation}{type=protocol,name=Activation,description={Final phase where an approved amendment becomes effective on mainnet}}
\newglossaryentry{quorum}{type=protocol,name=Quorum,description={Minimum participation threshold required for a governance vote phase to be valid}}
\newglossaryentry{supermajority}{type=protocol,name=Super-majority,description={Required affirmative percentage (e.g., 80\%) for an amendment vote to pass}}
\newglossaryentry{enshrined}{type=protocol,name=Enshrined,description={Implemented directly inside the protocol (not an external smart contract layer)}}
\newglossaryentry{smartcontract}{type=protocol,name=Smart Contract,description={User-deployed deterministic code executed by the protocol; distinct from enshrined logic}}
\newglossaryentry{onchain}{type=protocol,name=On-chain,description={Data/actions recorded directly in protocol state}}
\newglossaryentry{offchain}{type=protocol,name=Off-chain,description={Processes or data maintained externally (indexers, custody systems, reporting pipelines)}}
\newglossaryentry{layer2}{type=protocol,name=Layer-2,description={Secondary execution environment (e.g., rollup) anchored to Layer-1 for security/data availability}}
\newglossaryentry{genesis}{type=protocol,name=Genesis,description={Inaugural block from which chain history starts}}
\newglossaryentry{mainnet}{type=protocol,name=Mainnet,description={Production Tezos network}}
\newglossaryentry{testnet}{type=protocol,name=Testnet,description={Public network for staging, testing, and validation before mainnet activation}}
\newglossaryentry{rpc}{type=protocol,name=RPC,description={Remote Procedure Call interface exposing protocol state and helper queries}}
\newglossaryentry{ledger}{type=protocol,name=Ledger,description={Canonical record of account balances and token supplies}}
\newglossaryentry{celestezprotocol}{type=mechanics,name=\ProjectName\ protocol,description={Enshrined liquid staking mechanism managing aggregated stake, issuance, redemption, and validator allocation; issues the canonical liquid staking token \TokenTicker{}}}
\newglossaryentry{liquidstaking}{type=mechanics,name=Liquid Staking,description={Model issuing a transferable token representing staked exposure while underlying tez remain bonded}}
\newglossaryentry{directstaking}{type=mechanics,name=Direct Staking,description={Locking tez directly with a validator; funds are illiquid until unbonding completes}}
\newglossaryentry{fungibletoken}{type=mechanics,name=Fungible Token,description={Interchangeable token type; \TokenTicker{} follows the FA2.1 fungible interface}}
\newglossaryentry{clstz}{type=mechanics,name=\TokenTicker{},description={Canonical liquid staking token representing a proportional claim on aggregated tez}}
\newglossaryentry{tez}{type=mechanics,name=Tez,description={Native asset (XTZ) used for fees, staking, and governance}}
\newglossaryentry{mint}{type=mechanics,name=Mint,description={Creation of new \TokenTicker{} units upon deposit at prevailing exchange rate}}
\newglossaryentry{burn}{type=mechanics,name=Burn/Burning,description={Destruction of \TokenTicker{} units upon redemption request; corresponding tez move to the Frozen Ledger}}
\newglossaryentry{escrow}{type=mechanics,name=Escrow,description={Non-interest-bearing holding state for tez moved from the frozen ledger at finalization, pending user claim}}
\newglossaryentry{redemptionticket}{type=mechanics,name=Redemption Ticket,description={Internal non-transferable record capturing the claim on frozen tez and the target finalization cycle}}
\newglossaryentry{adaptiveslashing}{type=mechanics,name=Adaptive Slashing,description={Proportional penalty model where slashing severity scales with fraction of double attestations/bakings}}
\newglossaryentry{redemption}{type=mechanics,name=Redemption/Unstake,description={Two-step exit: burn tokens then wait; finalisation releases frozen tez after governance-set unbonding period}}
\newglossaryentry{unbonding}{type=mechanics,name=Unbonding,description={Waiting interval between burn request and finalisation while underlying tez remains slashable (reflected via $R_t$)}}
\newglossaryentry{frozen}{type=mechanics,name=Frozen,description={State where validator-bonded tez cannot transfer and remain subject to slashing risk}}
\newglossaryentry{slashing}{type=mechanics,name=Slashing,description={Penalty reducing a validator's frozen stake for consensus violations; reflected as lower exchange rate}}
\newglossaryentry{allocation}{type=mechanics,name=Allocation,description={Distribution of aggregated tez across eligible validators under caps and fee competitiveness}}
\newglossaryentry{cap}{type=mechanics,name=Cap,description={Limit preventing excessive concentration of aggregated stake on a single validator}}
\newglossaryentry{fee}{type=mechanics,name=Fee,description={Validator-declared reward share influencing net return and allocation ordering}}
\newglossaryentry{selfbond}{type=mechanics,name=Self-bond,description={Validator's own tez staked; security deposit sizing permissible external allocations}}
\newglossaryentry{enshrinedrate}{type=mechanics,name=Enshrined Exchange Rate,description={Canonical rate $R_t=\frac{L_t}{S_t}$ (tez per token); $L_t$ excludes frozen/escrowed tez; $S_t$ is the circulating supply (burned units excluded)}}
\newglossaryentry{nav}{type=finance,name=NAV,description={Net Asset Value: Aggregate tez (or reference-currency) value of holdings: $H_t \times R_t$ (optionally $\times$ FX$_t$)}}
\newglossaryentry{inav}{type=finance,name=iNAV,description={Intraday indicative NAV estimate published between formal valuation points}}
\newglossaryentry{fx}{type=finance,name=FX,description={Foreign exchange rate mapping tez value into another currency}}
\newglossaryentry{exchangerate}{type=finance,name=Exchange Rate,description={Tez per \TokenTicker{} internally; externally currency conversion multiplier (FX)}}
\newglossaryentry{intrinsicvalue}{type=finance,name=Intrinsic Value,description={Fundamental tez-denominated value per token: $R_t$ (plus FX if cross-currency)}}
\newglossaryentry{premiumdiscount}{type=finance,name=Premium/Discount,description={Market price deviation above/below intrinsic value ($R_t$ or $R_t \times$ FX)}}
\newglossaryentry{arbitrage}{type=finance,name=Arbitrage,description={Trade exploiting premium/discount by mint-and-sell or buy-and-redeem flows to restore alignment}}
\newglossaryentry{accrual}{type=finance,name=Accrual,description={Economic return embedded in rising $R_t$ without discrete distributions}}
\newglossaryentry{distribution}{type=finance,name=Distribution,description={Periodic payout (absent in \ProjectName; replaced by continuous accrual in $R_t$)}}
\newglossaryentry{costbasis}{type=finance,name=Cost Basis,description={Entry exchange rate (and FX if applicable) used to compute realised gain/loss at exit}}
\newglossaryentry{realization}{type=finance,name=Realization,description={Conversion of accrued appreciation into a taxable/disclosed event via redemption or sale}}
\newglossaryentry{marktomarket}{type=finance,name=Mark-to-Market,description={Period-end valuation using current $R_t$ (and FX) even absent redemption}}
\newglossaryentry{yield}{type=finance,name=Yield,description={Effective annualised percentage return implied by $R_t$ appreciation versus starting period}}
\newglossaryentry{exposure}{type=finance,name=Exposure,description={Economic stake: quantity of \TokenTicker{} $\times R_t$ held}}
\newglossaryentry{custody}{type=finance,name=Custody/Custodian,description={Safekeeping infrastructure holding tokens; segregated from protocol logic}}
\newglossaryentry{liquidity}{type=finance,name=Liquidity,description={Ease of converting \TokenTicker{} to tez (redemption) or other assets (secondary trading)}}
\newglossaryentry{audittrail}{type=finance,name=Audit Trail,description={Chronological on-chain record of lifecycle and allocation events enabling verification}}
\newglossaryentry{aml}{type=regulatory,name=AML,description={Anti-Money Laundering rules targeting illicit flows}}
\newglossaryentry{kyc}{type=regulatory,name=KYC,description={Know Your Customer identity/due diligence procedures}}
\newglossaryentry{compliance}{type=regulatory,name=Compliance,description={Adherence to applicable legal, regulatory, and reporting frameworks}}
\newglossaryentry{beneficialownership}{type=regulatory,name=Beneficial Ownership,description={Identification of natural person ultimately controlling or benefiting from an asset}}
\newglossaryentry{soc}{type=regulatory,name=SOC,description={Service Organization Control audit reporting on internal controls}}
\newglossaryentry{segregated}{type=regulatory,name=Segregated,description={Separation of client assets from operational/proprietary holdings}}
\newglossaryentry{wrapper}{type=regulatory,name=Wrapper,description={External legal/product structure (e.g., ETF) encapsulating protocol exposure}}
\newglossaryentry{sanctions}{type=regulatory,name=Sanctions,description={Regulatory prohibitions on transacting with specified entities/jurisdictions}}
\newglossaryentry{api}{type=tech,name=API,description={Programmatic interface for integration or data access}}
\newglossaryentry{hash}{type=tech,name=Hash,plural=Hashes,description={Fixed-size cryptographic digest uniquely representing data}}
\newglossaryentry{multisig}{type=tech,name=Multisig,description={Multi-signature scheme requiring multiple keys to authorise an operation}}
\newglossaryentry{mpc}{type=tech,name=MPC,description={Multi-Party Computation technique enabling distributed key control without single-party exposure}}
\newglossaryentry{fallback}{type=tech,name=Fallback,description={Secondary data source/procedure when primary feed is unavailable/stale}}
\newglossaryentry{allowlist}{type=tech,name=Allowlist,description={Controlled list governing which addresses may interact with a restricted interface}}
\newglossaryentry{threshold}{type=tech,name=Threshold,description={Minimum quantitative requirement (performance or quorum) for eligibility/progression}}
\newglossaryentry{ema}{type=tech,name=EMA,description={Exponential Moving Average used to adapt quorum or participation parameters}}
\newglossaryentry{dex}{type=tech,name=DEX,description={Decentralised Exchange enabling peer-to-peer token swaps (e.g., \TokenTicker{}/XTZ) without central custody}}
\makeglossaries
\usepackage{authblk}
\title{\textbf{\ProjectName}\\[4pt]\large A Protocol-Native Liquid Staking Solution for Tezos}
\author[1]{Mathias Bourgoin}
\author[2]{Arthur Breitman}
\author[1]{Pierrick Couderc}
\author[1]{Zaynah Dargaye}
\author[1]{Diane Gallois-Wong}
\author[1]{Marina Polubelova}
\author[1]{Lucas Randazzo}
\author[1]{Julien Tesson}

\affil[1]{Nomadic Labs}
\affil[2]{Tezos Co-founder}

\date{Compiled on \DTMnow}

\begin{document}
\makeatletter
\begin{center}
	{\LogoGraphic}\\[0.5em]
	{\LARGE \@title}\\[0.6em]
	{\small \@author}\\[0.6em]
	{\small \@date}
\end{center}
\makeatother

\begin{abstract}
	\ProjectName\ (\TokenTicker{}) is an \emph{enshrined}, protocol-native mechanism designed to mitigate the centralization risks associated with liquid staking intermediaries. Intended to complement direct staking rather than replace it, \ProjectName\ provides a neutral, public alternative managed directly by the Tezos protocol. It allows any tez holder to participate in aggregated staking without reliance on third-party operators. \TokenTicker{} follows an accrual-based design: all slashing events and rewards are reflected in the token's exchange rate to tez, keeping balances fungible while exposing holders to the precise economics of staking. This approach ensures that liquid staking functions as fundamental network infrastructure—with deterministic lifecycle rules, transparent on-chain data, and governance anchored in the amendment process—rather than as a discretionary commercial product. This white paper summarises the motivation for enshrining liquid staking, the core mechanics, exchange-rate model, regulatory touchpoints, risk posture, and forward-looking roadmap for \ProjectName.
\end{abstract}
\setcounter{tocdepth}{1}
\tableofcontents

\medskip

\section{Overview}
\label{sec:overview}
\subsection{Executive Summary}\label{subsec:exec-summary}
\ProjectName{} (\TokenTicker{}) is an enshrined, protocol-native mechanism for \gls{tezosprotocol} that aggregates user stake into a unified system managed directly by the protocol. While direct staking remains the optimal method for network health and individual sovereignty, the emergence of centralized liquid staking intermediaries creates centralization risks. \ProjectName{} mitigates these risks by providing a permissionless, decentralized alternative: a protocol-level system that allows any \gls{tez} holder to participate in staking via a fungible claim, without relying on third-party \glspl{smartcontract} or \glspl{custody}.

\textbf{\TokenTicker{} is the Canonical Liquid Staking Token:}\label{def:clstz-definition} a protocol-issued \gls{fungibletoken} (implementing Tezos' \href{https://docs.tezos.com/architecture/tokens/FA2.1}{FA2.1 token standard}) that represents a proportional share of the aggregated stake. \ProjectName{} follows an \emph{accrual} model: rewards and slashing outcomes modify the exchange rate rather than adjusting wallet balances. The system is designed for neutrality and security:
\begin{itemize}
	\item Any tez account can deposit into the shared staking \gls{ledger} via the protocol entrypoints.
	\item Depositors receive \TokenTicker{} tokens that are fully fungible: each unit represents the same proportional share of the total staked amount.
	\item The quantity minted is determined by the prevailing exchange rate \ExchangeRateSymbol{} (where \ExchangeRateSymbol{} denotes the amount of tez backing a single unit of \TokenTicker{} at block $t$).
	\item To exit, a holder submits an unstake request, burning tokens immediately at the current exchange rate. After the standard unbonding period, they receive the frozen tez (less any slashing penalties accrued during the wait).
	\item Staking rewards increase the total staked balance and slashing events reduce it; both effects are expressed through \ExchangeRateSymbol{}, keeping token balances unchanged while reflecting the precise economics of the underlying stake.
\end{itemize}

\paragraph{Enshrined Interface.} A single protocol-native interface contract mediates all user interactions. It is enshrined (implemented inside the Tezos protocol) and canonical. Its token entrypoints use a standard fungible interface so existing wallets and exchanges can interact with \TokenTicker{} as a standard asset. The same contract manages redemption queues, enforces the unbonding period, and emits lifecycle events, eliminating administrator key risk.

\paragraph{About Tezos.} Tezos is a \gls{pos} Layer-1 blockchain with on-chain governance and a forkless self-amendment mechanism that has enabled 20 protocol upgrades since mainnet launch in 2018—without contentious chain splits. The governance pipeline lets protocol changes be proposed, voted on, and activated entirely in-protocol, keeping consensus aligned with stakeholder decisions. The reference implementation is written in OCaml—a language used in safety-critical and high-reliability software (e.g., at Airbus and Jane Street)—providing high-level safety features such as a strong static type system, algebraic data types, and automatic memory management. Tezos supports a rich smart-contract ecosystem and modern scaling features (including Layer-2 rollups and a data-availability layer), targets ~6-second blocks with near-instant two-block finality, and has experienced no protocol-level downtime or successful chain attacks since launch. \\
Throughout this document we use the neutral term \emph{\gls{validator}} for block producers/voters; in the Tezos ecosystem they are traditionally called ``bakers''.


\paragraph{Key properties.} The design establishes staking as a neutral, foundational component of the Tezos ecosystem.\vspace{0.25em}
\begin{itemize}
	\item \textbf{Transparent Exchange Rate:} Single on-chain metric $\ExchangeRate$ (tez per token) defined as $\LedgerBalance/\TokenSupply$, calculated at block level $t$. Components:\newline\hspace*{1em}$\LedgerBalance$ — total tez in the aggregated stake (contract storage / RPC).\newline\hspace*{1em}$\TokenSupply$ — circulating \TokenTicker{} units (fungible token ledger).\newline This quotient is the canonical measure of value.
	\item \textbf{Algorithmic Allocation:} Any eligible validator can register to receive stake allocations. Eligilibilty criteria are defined at the protocol level. The system automatically distributes stake between registered validators, promoting decentralization and preventing any single operator from capturing disproportionately more stake than achievable through standard delegation.
	\item \textbf{Core Protocol Security:} The entire system is built into the Tezos protocol itself, rather than being layered on top of it via complex smart contracts. This minimizes custom contract and custodial management risks, and ensures that any future upgrades are subject to the formal, community-driven Tezos amendment process.
	\item \textbf{Deterministic Lifecycle:} The system's architecture ensures a predictable operational lifecycle with straightforward accrual accounting and a complete, transparent audit trail.
\end{itemize}


\subsection{Mitigating Centralization Risks}
Liquid staking on proof-of-stake networks has typically been handled by third-party services. While functional, these models can concentrate control over the network's validation process and introduce unique operational risks. Liquid staking markets exhibit strong network effects driven by liquidity: the most liquid token becomes the most useful for DeFi and exit, attracting more users in a self-reinforcing loop. This creates a natural 'winner-take-all' dynamic. If a private intermediary captures this liquidity monopoly, they gain disproportionate control over the network's consensus. \ProjectName{} addresses this by providing a canonical, protocol-native alternative that removes operator discretion from the core functions of issuance and lifecycle management. By enshrining these mechanics, the protocol ensures that liquid staking does not become a vector for centralization.

\subsection{Key Benefits for Ecosystem Health}\label{para:key-properties}
\begin{itemize}
	\item \textbf{Verifiable Exchange Rate:} The tez-denominated value is computed from two on-chain data points: the total tez in the system (\LedgerBalance) and total \TokenTicker{} supply (\TokenSupply), determining \ExchangeRateSymbol{}).
	\item \textbf{Simplified Accounting:} Because rewards are reflected in the token's potential appreciation when denominated in tez rather than being paid out, the system avoids the complexity of rebasing mechanics.
	\item \textbf{On-chain Transparency:} All lifecycle actions (deposit, redemption) and key protocol events are recorded on-chain, providing an end-to-end audit trail.
	\item \textbf{Clear Separation of Duties:} The protocol handles the core mechanics of staking. External entities (such as custodians or wallets) can interact with the system without needing special permissions.
	\item \textbf{Predictable Governance:} Changes to the system's parameters must go through the Tezos amendment process, a formal on-chain voting cycle with fixed-duration phases and public validator ballots. This ensures that the rules of the system cannot be changed arbitrarily by a centralized party (see Section~\ref{sec:governance}).
\end{itemize}

\begin{center}
	\fbox{%
		\begin{minipage}{0.9\linewidth}
			\textbf{Protocol Guarantees (Summary).} Lifecycle rules are deterministic and embedded in the Tezos protocol; there are no privileged administrator keys, and any parameter change (caps, eligibility, fees, redemption policy) must progress through the amendment pipeline described in Section~\ref{sec:governance}. Later sections reference this summary rather than restating it verbatim.
		\end{minipage}%
	}
\end{center}

\subsection{Comparison with Centralized Intermediaries}
To orient readers, Table~\ref{tab:comparison} contrasts a protocol-enshrined design with operator-run liquid staking services across governance, validator diversity, fee discovery, reward handling, slashing treatment, data access, and upgrade process. In short, \ProjectName{} minimizes discretionary control and exposes a simpler, auditable surface (two public scalars for the exchange rate), whereas private intermediaries centralize decisions. Even when governed by a DAO, these intermediaries introduce an additional layer of trust distinct from the protocol. DAO incentives may diverge from the underlying chain's health, and their governance is prone to capture—where large incumbents can block new validators to protect margins. \ProjectName{} removes this layer, anchoring governance directly to the Tezos amendment process.

\Needspace{12\baselineskip}
\begin{table}[H]
	\centering
	\renewcommand{\arraystretch}{1.05}
	\begin{tabularx}{0.98\linewidth}{p{3.1cm} X X}
		\toprule
		\textbf{Dimension}   & \textbf{\ProjectName{}}                                                                                                          & \textbf{Centralized Intermediaries}                                                                      \\
		\midrule
		Control              & Governed by the Tezos protocol and its formal amendment process.                                                                 & Typically governed by a single corporation or a small multisig group.                                    \\
		Fee Structure        & Fees are set by validators in an open protocol mechanism.                                                                        & Fees are set by the operator, often with limited transparency.                                           \\
		Validator Set        & Designed to be diverse, with stake caps preventing concentration.                                                                & Often concentrated among a few large, proprietary operators.                                             \\
		Reward Handling      & Rewards accrue automatically, reflected in the token's exchange rate.                                                            & Rewards are frequently distributed as separate transactions to each user.                                \\
		Slashing Risk        & Losses are shared pro\mbox{-}rata across the entire system via a rate adjustment. The faulty validator is automatically removed. & Remediation is handled on a case-by-case basis and may depend on insurance.                              \\
		Data Access          & All data is public on-chain. Exchange rate is derived from two simple values.                                                    & Data is often available only through proprietary dashboards or \glspl{api}.                              \\
		Upgrades             & Follow the predictable, multi-stage Tezos amendment process.                                                                     & Timelines and scope are at the operator's discretion.                                                    \\
		Governance Alignment & Fully Aligned: Changes require a super-majority of the entire network via the L1 amendment process.                              & Misaligned Layer: Governed by a specific token/DAO, creating a separate interest group prone to capture. \\
		\bottomrule
	\end{tabularx}
	\caption{Contrast between \ProjectName{} and operator-run liquid staking services.}
	\label{tab:comparison}
\end{table}

\subsection{Lifecycle Walkthrough}
\label{sec:lifecycle}
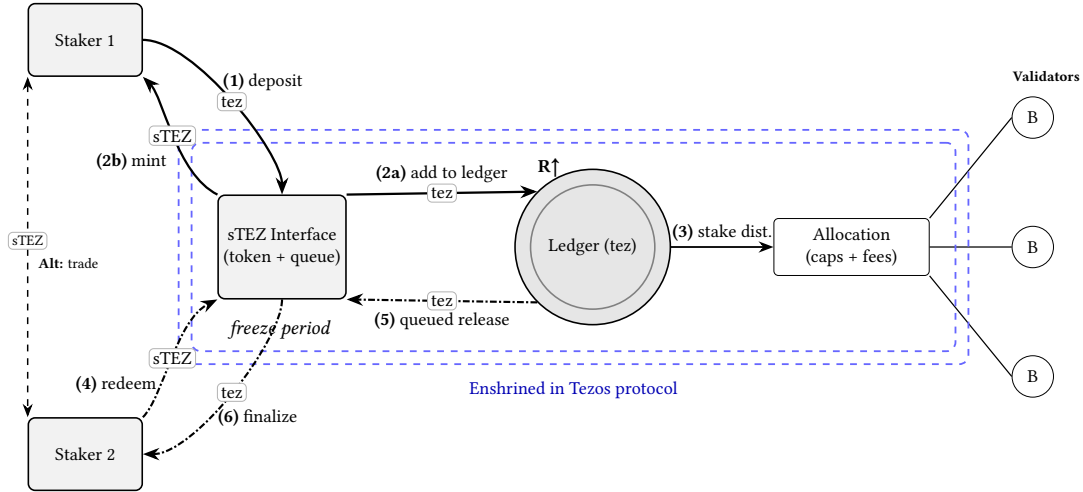
\begin{figure}[!htbp]
	\centering
\resizebox{0.95\linewidth}{!}{\begin{tikzpicture}[font=\small,>=Stealth,inner sep=3pt,scale=0.8,transform shape]
  \tikzset{
    procBox/.style={rounded corners=3pt,draw=black,fill=black!5,line width=0.7pt,inner sep=3pt,minimum width=22mm,minimum height=14mm,align=center},
    ledgerCircle/.style={circle,draw=black,fill=black!10,line width=0.7pt,minimum size=30mm,inner sep=1.6pt},
    validatorNode/.style={circle,draw=black,fill=white,minimum size=8mm,inner sep=0pt},
    flowMain/.style={->,line width=1pt},
    flowRedeem/.style={->,line width=0.95pt,densely dashdotted},
    flowTrade/.style={<->,dashed,line width=0.65pt},
    flowAlloc/.style={->,line width=0.8pt},
    thinLink/.style={-,line width=0.55pt}
  }
  \node[ledgerCircle] (LEDGER) at (0,0) {Ledger (tez)};
  \node[procBox,minimum height=20mm] (PROTOCOL) at ($(LEDGER)+(-60mm,0)$) {\TokenTicker{} Interface\\(token + queue)};
  \node[draw=black,rounded corners=2pt,minimum width=30mm,minimum height=11mm,align=center,fill=white] (ALLOC) at ($(LEDGER)+(50mm,0)$) {Allocation\\(caps + fees)};
  \node[procBox] (STAKER) at ($(PROTOCOL)+(-38mm,40mm)$) {Staker 1};
  \node[procBox] (STAKER2) at ($(PROTOCOL)+(-38mm,-40mm)$) {Staker 2};
  \node[validatorNode] (B2) at ($(ALLOC)+(35mm,25mm)$) {B};
  \node[validatorNode] (B1) at ($(ALLOC)+(35mm,0)$) {B};
  \node[validatorNode] (B3) at ($(ALLOC)+(35mm,-25mm)$) {B};
  \draw[thinLink] (ALLOC.east) -- (B1.west);
  \draw[thinLink] (ALLOC.north east) -- (B2.west);
  \draw[thinLink] (ALLOC.south east) -- (B3.west);
  \node[font=\scriptsize,anchor=west] at ($(B2.north)+(-5mm,+4mm)$) {\textbf{Validators}};
  \node[draw=blue!60,thick,dashed,rounded corners=4pt,inner sep=4mm,fit={(PROTOCOL) (LEDGER) (ALLOC)}] (ENSHRINE_INNER) {};
  \node[draw=blue!60,thick,dashed,rounded corners=4pt,inner sep=6mm,fit={(PROTOCOL) (LEDGER) (ALLOC)}] (ENSHRINE_OUTER) {};
  \node[font=\small,anchor=north,text=blue!70!black,inner sep=1mm] at ($(ENSHRINE_OUTER.south)+(0,-2mm)$) {Enshrined in Tezos protocol};
  \draw[black!50,line width=0.7pt] (LEDGER) circle (12mm);
  \node[font=\small,anchor=south west] at ($(LEDGER.north west)+(-1mm,1.8mm)$) {\textbf{R↑}};
  \draw[flowMain] (STAKER.east) .. controls ($(STAKER.east)+(10mm,0)$) and ($(PROTOCOL.north)+(0,8mm)$) .. coordinate[pos=0.5] (A1) (PROTOCOL.north);
  \node[above, font=\small, xshift=6mm, yshift=1mm] at (A1) {\textbf{(1)} deposit};
  \node[draw=black!30,fill=white,rounded corners=2pt,inner sep=2pt,font=\small] at (A1) {tez};
  \draw[flowMain] (PROTOCOL.north east) -- coordinate[pos=0.5] (A2A) (LEDGER.north west);
  \node[above,font=\small, yshift=1mm] at (A2A) {\textbf{(2a)} add to ledger};
  \node[draw=black!30,fill=white,rounded corners=2pt,inner sep=2pt,font=\small] at (A2A) {tez};
  \draw[flowMain] (PROTOCOL.north west) .. controls ($(PROTOCOL.north west)+(-8mm,4mm)$) and ($(STAKER.south east)+(4mm,-4mm)$) .. coordinate[pos=0.5] (A2B) (STAKER.south east);
  \node[left,font=\small, yshift=-5mm] at (A2B) {\textbf{(2b)} mint};
  \node[draw=black!30,fill=white,rounded corners=2pt,inner sep=2pt,font=\small] at (A2B) {\TokenTicker{}};
  \draw[flowTrade] (STAKER.south west) -- coordinate[pos=0.5] (AT) (STAKER2.north west);
  \node[right,font=\scriptsize,xshift=1mm, yshift=-3mm] at (AT) {\textbf{Alt:} trade};
  \node[draw=black!30,fill=white,rounded corners=2pt,inner sep=2pt,font=\scriptsize,yshift=1.2mm] at (AT) {\TokenTicker{}};
  \draw[flowAlloc] (LEDGER.east) -- node[above,font=\small]{\textbf{(3)} stake dist.} (ALLOC.west);
  \draw[flowRedeem] (STAKER2.north east) .. controls ($(STAKER2.north east)+(4mm,4mm)$) and ($(PROTOCOL.south west)+(-8mm,-4mm)$) .. coordinate[pos=0.5] (A4) (PROTOCOL.south west);
  \node[left,font=\small, xshift=-2mm, yshift=-5mm] at (A4) {\textbf{(4)} redeem};
  \node[draw=black!30,fill=white,rounded corners=2pt,inner sep=2pt,font=\small] at (A4) {\TokenTicker{}};
  \draw[flowRedeem] (LEDGER.south west) -- coordinate[pos=0.5] (A5) (PROTOCOL.south east);
  \node[below,font=\small, yshift=-1mm] at (A5) {\textbf{(5)} queued release};
  \node[draw=black!30,fill=white,rounded corners=2pt,inner sep=2pt,font=\small] at (A5) {tez};
  \draw[flowRedeem] (PROTOCOL.south) .. controls ($(PROTOCOL.south)+(0,-8mm)$) and ($(STAKER2.east)+(10mm,0)$) .. node[below,pos=0.1,font=\small,font=\itshape]{freeze period} coordinate[pos=0.5] (A6) (STAKER2.east);
  \node[below,font=\small, xshift=5mm, yshift=-2mm] at (A6) {\textbf{(6)} finalize};
  \node[draw=black!30,fill=white,rounded corners=2pt,inner sep=2pt,font=\small] at (A6) {tez};
\end{tikzpicture}}
	\caption{Lifecycle overview for \ProjectName{}, including the freeze delay that protects against post-denunciation slashing before tez are released.}
	\label{fig:lifecycle-overview}
\end{figure}

\noindent\textit{Protocol lifecycle at a glance.} Figure~\ref{fig:lifecycle-overview} summarises the flow: (1) users deposit tez; (2a) tez is added to the aggregated stake; (2b) \TokenTicker{} is minted to the staker; (3) stake is allocated to validators under caps and protocol criteria; (4) a redemption request burns \TokenTicker{}; (5) tez is queued to the interface contract; (6) finalization releases tez after the governance-set unbonding period.
\Needspace{15\baselineskip}

To complement the figure, the walkthrough below highlights how deposits, transfers, and redemptions unfold for participants interacting with the \ProjectName{} protocol:

\noindent\textbf{Entry (Deposit and Mint).} Staker~1 deposits tez to the interface contract (step~1). The \ProjectName{} protocol adds the deposited tez to the aggregated stake (step~2a) and mints new \TokenTicker{} tokens (step~2b), which Staker~1 receives. These tokens represent pro-rata ownership of the staked funds. Subsequent allocation cycles (step~3) distribute the staked tez across eligible validators under capacity caps and fee constraints. Staking rewards accrue in the system, increasing the \TokenTicker{}/tez exchange rate ($\ExchangeRateSymbol\uparrow$).

\noindent\textbf{Trade (Secondary Transfer).} Staker~1 may trade their \TokenTicker{} tokens to Staker~2 via a decentralised exchange (DEX) or direct transfer (Alt step), providing liquidity without triggering protocol-level redemption or freeze periods.

\noindent\textbf{Exit (Redeem and Release).} Staker~2, now holding \TokenTicker{}, initiates a redemption by submitting a request to the \ProjectName{} protocol (step~4), burning their tokens. The \ProjectName{} protocol queues the release (see Redemption Queue Semantics in Section~\ref{sec:mechanics}), transferring the corresponding tez from the aggregated stake to the interface contract (step~5). The same contract enforces the unbonding period, during which the tez remain frozen. After the unbonding period expires, the contract finalizes the release (step~6), transferring the tez directly to Staker~2's wallet. Economic gain or loss is realised at redemption based on the exchange rate at that time $t$.

\section{Mechanics}
\label{sec:mechanics}

This section defines the protocol-level state variables, transitions, and invariants that govern the operation of the enshrined liquid staking mechanism. All terms used here are defined either inline or in the Glossary (Section~\ref{sec:glossary}). All transitions are deterministic and require no \gls{offchain} coordination, no operator, and no fund-like structure.

Throughout this section:
\begin{itemize}
  \item $t \in \mathbb{N}$ denotes a block level.
  \item $c(t) \in \mathbb{N}$ denotes the cycle index containing block $t$.
  \item $L_t \in \mathbb{R}_{\ge 0}$ denotes the total amount of \gls{tez} tracked in the system-level staking \gls{ledger} at block $t$.
  \item $S_t \in \mathbb{R}_{\ge 0}$ denotes the total supply of \TokenTicker{} tokens at block $t$.
  \item $R_t = L_t / S_t$ denotes the exchange rate at block $t$, expressed in tez per unit token (when $S_t > 0$).
\end{itemize}

By construction, $L_t$ and $S_t$ evolve only through protocol-defined transitions described below.

\subsection{System State}
At any block $t$, the mechanism maintains the following state variables:
\begin{enumerate}
  \item \textbf{Staking ledger $L_t$}. The quantity of tez actively staked through the mechanism and tracked by the protocol.
  \item \textbf{Token supply $S_t$}. The number of \TokenTicker{} units outstanding.
  \item \textbf{Validator registry $V_t$}. A set of \glspl{validator} eligible to receive stake assignments at cycle $c(t)$. For each validator $v \in V_t$, the protocol records:
  \begin{itemize}
    \item a fee parameter $f(v)$,
    \item a maximum capacity $cap(v)$ (maximum tez assignable),
    \item performance counters (for slashing and reward computation).
  \end{itemize}
  \item \textbf{Frozen ledger $F_t$}. A mapping from cycle indices to tez amounts that are unbonding. These funds do not accrue rewards but remain subject to slashing.
  \item \textbf{Finalizable ledger $E_t$}. A single balance of tez that has completed unbonding and is available for withdrawal.
\end{enumerate}
These variables form the minimal state representation needed to describe all transitions.

\paragraph{Protocol Timing}
Within a block, user operations (deposits, redemptions) apply first, so the exchange rate reflects those updates for that block. At the cycle boundary, rewards and any slashing are applied together to the staking ledger; immediately after, the protocol computes stake allocation for the next cycle. This separates block-level state changes from cycle-level transitions while keeping $R_t$ consistent with the latest operations and end-of-cycle adjustments.

\paragraph{Observable Data Surface}
Every state change is captured in deterministic events:
\begin{itemize}
  \item \textbf{Deposit:} depositor address, tez credited, \TokenTicker{} minted.
  \item \textbf{Redemption requested:} requester, tokens burned, tez frozen, eligibility timestamp.
  \item \textbf{Redemption finalized:} which ticket was finalized and tez released.
  \item \textbf{Stake allocation:} cycle, validator, tez assigned, effective fee, cap flags.
  \item \textbf{Slashing:} cycle(s) affected, validator identity, percentage penalty, resulting change in \LedgerBalance{} and frozen buckets.
\end{itemize}
Read-only interfaces expose only the scalars needed for reconciliation: \LedgerBalance{}, \TokenSupply{}, \ExchangeRateSymbol{}, per-validator assignments and fees, the frozen ledger (with freeze metadata), and eligibility status. This minimal surface keeps the protocol auditable and allows custodians and observers to reconstruct history without privileged access.

\subsection{Deposit and Minting}
A deposit is a transition that increases $L_t$ and $S_t$ synchronously. At block $t$, an account may invoke the deposit entrypoint with amount $\Delta \in \mathbb{R}_{\ge 0}$ tez. The protocol executes:
\begin{enumerate}
  \item \textbf{Minting:} $u = \frac{\Delta}{R_t}$ units of \TokenTicker{} are minted (assuming $S_t > 0$; the genesis case is handled separately by initialization).
  \item \textbf{Ledger update:} $L_t := L_t + \Delta,\qquad S_t := S_t + u.$
  \item \textbf{Assignment eligibility:} the deposited tez becomes eligible for validator assignment at the boundary between $c(t)$ and $c(t)+1$.
\end{enumerate}
This transition preserves the exchange rate $R_t = L_t / S_t$. No intermediate entity holds $\Delta$; the transition updates the protocol's internal staking ledger directly.

\paragraph{User View (Deposit)}
Submit tez to the interface contract (directly or via wallet integrations). The contract mints \TokenTicker{} immediately, crediting the user’s fungible token balance. The amount minted equals \emph{tez deposited}~$\div R_t$, where the exchange rate $R_t$ is published on-chain. Once rewards have accumulated and $R_t>1$, each mutez (1 tez = 1{,}000{,}000 mutez) mints fewer than one token—there is no fixed one-to-one peg after \gls{genesis}.

\subsection{Validator Assignment}
Before the first block of cycle $k$ (i.e., at the last block of cycle $k\!-\!1$), the protocol computes a stake-distribution function $A_k : V_{t_k} \longrightarrow \mathbb{R}_{\ge 0}$ that governs validator rights in cycle $k + \texttt{consensus\_rights\_delay}$, where $\texttt{consensus\_rights\_delay}$ is a protocol constant (currently two full cycles). This computation runs alongside other protocol rights assignments at the cycle boundary and is subject to:
\begin{description}
  \item[Capacity constraint:] For all $v \in V_{t_k}$, $A_k(v) \le cap(v).$
  \item[Consistency constraint:] $\sum_{v \in V_{t_k}} A_k(v) = L_{t_k}.$
  \item[Eligibility constraint:] Only validators satisfying the objective criteria (e.g., minimum self-bond, clean recent slashing history) are included in $V_{t_k}$.
\end{description}
The assignment is deterministic, depends solely on on-chain data, and does not involve delegation, off-chain aggregation, or operator-level decisions.

\paragraph{User View (Validator)}
Validators interact with the system through a separate, permissionless flow. They register by declaring a fee and capacity. Consensus rights are first assigned at the very next cycle boundary after registration, but those rights take effect in the future cycle offset by the protocol’s consensus rights delay (currently two full cycles). With each cycle, the protocol ranks eligible validators by effective fee, respects per-validator caps, and assigns stake. Operators can monitor allocations, caps, and performance signals through the same read-only surfaces exposed to stakers.

\subsection{Reward Accrual}
Protocol rewards (baking, attestations, data-availability) are credited as blocks land according to the protocol’s reward schedule; when credited they increase $L_t$ (with $S_t$ unchanged), so $R_t$ can evolve within a cycle. Slashing reduces $L_t$ when applied (see below). At the end of block $t$, let $\Delta_t$ be the net block accrual that becomes available at the block's application. For the next block of cycle $t+1$:
\[
  L_{t+1} := L_{t} + \Delta_t,\qquad S_{t+1} := S_{t}.
\]
Consequently: $R_{t+1} = \frac{L_{t} + \Delta_t}{S_t}.$
Rewards do not modify $S_t$ and are never distributed as separate payments; they appear as changes to $L_t$ through protocol rules.

\subsection{Redemption and Settlement}
\paragraph{Immediate Burn (Request)}
At block $t$, a user requests to redeem $u$ tokens. The protocol \textbf{instantly} burns $u$ from $S_t$ and calculates the tez value $v = u \cdot R_t$.
It moves $v$ from the Staking Ledger ($L_t$) to the Frozen Ledger for the current cycle ($F_t[c(t)]$):
\[
  S_t := S_t - u
\]
\[
  L_t := L_t - v
\]
\[
  F_t[c(t)] := F_t[c(t)] + v
\]
\emph{Effect:} The user stops earning rewards immediately. The exchange rate $R_t$ remains constant for remaining stakers.

\paragraph{Unbonding Period (Slashing Risk)}
Funds in $F_t$ remain frozen for the governance-defined unbonding period. During this time:
\begin{itemize}
  \item \textbf{No Rewards:} Since the funds are not in $L_t$, they do not receive staking rewards.
  \item \textbf{Slashing Exposure:} If a slashing event reduces the Staking Ledger $L_t$ by a percentage $p$, the same percentage $p$ is deducted from all active buckets in the Frozen Ledger $F_t$.
  \item \textbf{Outstanding rights:} Frozen tez continues to back any consensus rights already issued for the relevant horizon; freezing prevents withdrawal until those rights have expired or been exercised.
\end{itemize}

\paragraph{Finalization (Cycle End)}
At the cycle boundary where a bucket $F_t[k]$ matures, the remaining balance is moved to the Finalizable Ledger:
\[
  E_t := E_t + F_t[k]
\]
\[
  F_t[k] := 0
\]
Users can then permissionlessly withdraw their share from $E_t$.

\paragraph{User View (Unstake)}
Burn the desired amount of \TokenTicker{} to receive a claim on frozen tez. The wallet shows the claim with its projected unlock cycle. After the unbonding period expires, the user—or any helper—calls finalize. The contract transfers the finalized tez straight back to the original requester. The claim is an internal record: it cannot be traded, reassigned, or otherwise exchanged during the wait.

\subsection{Slashing and Penalties}
If a validator $v \in V_t$ commits a slashable offense, the protocol computes a penalty $p(v) \ge 0$. At the block when the penalty is applied:
\[
  L_t := L_t - p(v),\qquad S_t := S_t.
\]
Thus, the exchange rate adjusts as: $R_t := \frac{L_t - p(v)}{S_t}.$ Any denunciation rewards credited as part of the slashing process are incorporated into subsequent $\Delta_k$ at cycle end. The mechanism remains well-defined even under partial or full validator failures, with no dependence on operator-managed funds.

\paragraph{Adaptive Slashing Context}
Tezos uses \gls{adaptiveslashing} for double attestations: the penalty scales with the fraction of attesters that double-sign at a level (small when isolated, larger when many collude). Double baking (two distinct blocks at the same level) follows a fixed baking-penalty schedule. Denunciations can be included during the faulty block’s cycle and the following one; penalties execute at the end of that window against the validator’s frozen stake (self-bond plus assigned stake). For \ProjectName{}, slashing burns tez at the protocol level. \LedgerBalance{} decreases, while \TokenSupply{} remains constant—no \TokenTicker{} is minted or burned. The result is a discrete drop in $\ExchangeRateSymbol$, so losses are borne pro-rata via a lower exchange rate per token. Slashed validators are automatically removed from the eligible set and must satisfy the attestation and history screens before being reconsidered.

\subsection{Invariants}
For all blocks $t$, the following invariants hold:
\begin{enumerate}
  \item \textbf{Zero-supply floor:} If $S_t = 0$, then $L_t = 0$ and we define $R_t := 1$ as a neutral quote (avoids division by zero while keeping the system well-posed at genesis or after full redemption).
  \item \textbf{Exchange-rate definition:} $R_t = \frac{L_t}{S_t} \quad \text{whenever } S_t > 0.$
  \item \textbf{Supply conservation (mint/burn only):} all increases to $S_t$ arise from deposit transitions, and all decreases arise from redemption transitions.
  \item \textbf{Ledger decomposition:} the staking ledger can be decomposed as $L_t = L_t^{\mathrm{assigned}} + L_t^{\mathrm{unassigned}},$ where $L_t^{\mathrm{assigned}}$ is the tez currently assigned to validators and $L_t^{\mathrm{unassigned}}$ is unassigned but still eligible stake. Note that $F_t$ and $E_t$ are tracked separately from $L_t$.
  \item \textbf{Deterministic evolution:} given $\sigma_t$ and the block contents (operations, slashing events, cycle boundaries), the next state $\sigma_{t+1}$ is uniquely determined by the protocol rules. No off-chain service, operator, or aggregated account is required.
\end{enumerate}

\section{Valuation and Methodology}
\label{sec:valuation}

\subsection{Indicative Net Asset Value}
Using the exchange rate defined in Section~\ref{sec:overview}, for a holder with $H_t$ tokens the canonical portfolio value is expressed directly in tez as $H_t \times \ExchangeRateSymbol$. Many reporting regimes additionally require fiat or reference-currency figures; in those cases the on-chain rate is paired with an external FX rate $\FXRate$:
\[
        \text{IndicativeNAV}_t = H_t \times \ExchangeRateSymbol \times \FXRate.
\]
The optional conversion requires only two on-chain scalars (\LedgerBalance{} and \TokenSupply{}) and a single external exchange-rate input. Funds can reuse the same expression to publish daily \glspl{nav} or intraday indicative values whenever cross-currency disclosure is needed.

\subsection{Data Inputs and Reconciliation}
Valuation depends on a minimal data set:
\begin{itemize}
    \item \textbf{Ledger balance (\LedgerBalance{}):} Queried from contract storage or an RPC view.
    \item \textbf{Token supply (\TokenSupply{}):} Reported by the interface contract’s fungible token ledger.
    \item \textbf{Event deltas:} Reward accrual and slashing amounts emitted as structured events for attribution.
    \item \textbf{Reference-currency FX (\FXRate):} Optional external FX rate feed (market price) when non-tez reporting is required.
\end{itemize}

Because both \LedgerBalance{} and \TokenSupply{} are on-chain, \ExchangeRateSymbol{} can be recomputed at block, intra-day, or cycle cadence. Timing effects follow Tezos primitives: deposits and redemption burns land immediately; block-level rewards accrue as blocks land; deferred reward components settle on the cycle schedule defined by adaptive issuance; denunciations can be included over the bounded slashing window before penalties execute at cycle end. Operators reconciling a window $[t_0, t_1]$ should attribute $\Delta R = R_{t_1}-R_{t_0}$ to (i) net rewards, (ii) net slashing losses, and (iii) deposits or redemptions straddling the cut-off. A tolerance band (for example 5 basis points) is typically sufficient to flag anomalies.

\subsection{NAV Reporting and Market Alignment}
Observers derive value (net asset value) packs by reading \LedgerBalance{} and \TokenSupply{} to compute \ExchangeRateSymbol{} in tez. When publications quote values in another currency, they multiply by the external FX rate \FXRate{}. Extraordinary events (e.g., a slash) warrant an accompanying notice referencing the relevant on-chain event but need no protocol action. Secondary-market pricing remains anchored by primary flows: if a market in the reference currency trades above $\ExchangeRateSymbol \times \FXRate$, mint-and-sell arbitrage restores parity; if it falls below, buy-and-redeem flows achieve the inverse. 
\section{Compliance and Integration}
\label{sec:regulatory}

This section explains how the protocol’s mechanics map to common regulatory, custody, and audit expectations. Where possible, information is sourced directly on-chain; \glspl{wrapper} (funds, brokers, custodians) handle their own \gls{compliance} layers \gls{offchain}.

\subsection{Neutral Protocol Basis}
The protocol-enshrined implementation eliminates discretionary administrator keys: deposits, redemptions, minting, and allocation all follow deterministic rules embedded in protocol code. This segregation enables regulated wrappers to layer compliance controls without relying on off-chain service providers for critical operations. 

\subsection{Deterministic Valuation and Disclosure}
Independent replication of $R_t$ from the on-chain balance/supply pair (\LedgerBalance{}, \TokenSupply{}) supports block-level computation, intraday indicative values, and (where applicable) daily NAV production for operational reporting. Slashing and allocation events are recorded as on-chain, structured logs emitted by the protocol. These can be consumed directly for audit trails, while any narrative disclosures (e.g., board memos) are compiled off-chain by the wrapper.

\subsection{Custody, Reporting, and Audit Workflows}
\begin{itemize}
  \item \textbf{Custody:} Custodians hold \TokenTicker{} using standard fungible token wallets; no separate reward inflows simplifies reconciliation and audit trails.
  \item \textbf{Valuation:} External systems read $R_t$ from the protocol view to compute tez-denominated value. When filings must cite another currency, they pair $R_t$ with an external tez exchange rate.
  \item \textbf{Recordkeeping:} The protocol emits per-request identifiers for redemption tickets on-chain. Off-chain systems (e.g., transfer agent ledgers) may mirror these IDs for convenience, but the source of truth for lifecycle events (deposits, redemptions, reward accrual, slashing) remains on-chain.
\end{itemize}

\subsection{Integrity Controls}
Stake caps, enforced eligibility criteria, and automated exclusion of underperforming validators provide quantifiable guardrails against concentration. In brief: "stake caps" limit how much of the system can be allocated to any one validator (relative to their self-bond and global caps), and "eligibility" requires a clean recent history (no slash), declared fee, and declared capacity. See Section~2.4 (Validator Registration and Stake Allocation) for mechanics. These controls map to standard risk disclosures (e.g., limits on validator exposure, triggers for governance actions) and support board oversight.

\subsection{Upgrade and Notice Process}
Tezos amendments progress through proposal, exploration, testing, promotion, and activation periods (see \href{https://docs.tezos.com/architecture/governance}{governance architecture}), with fixed durations that collectively exceed several weeks. Governance documentation will include notice templates so operators can communicate pending parameter changes, update disclosures, or adjust operational runbooks before activation.

\subsection{Compliance Layering and Responsibility Boundaries}
The Tezos protocol remains permissionless. Compliance obligations (\gls{kyc}/\gls{aml} screening, \gls{sanctions} checks, \gls{beneficialownership} reporting) rest with wrappers such as custodians, issuers, or broker-dealers. The deterministic lifecycle reduces the scope of bespoke policies: wrappers must police primary-market inflows, while secondary trading can lean on existing exchange frameworks. 

\section{Risk and Security Overview}
\label{sec:risk}

\subsection{A Framework for Risk}
The security of \ProjectName{} is assessed across several key domains: the integrity of the protocol's implementation, the behavior of network validators, the potential for economic manipulation, market liquidity, governance security, data availability, and the evolving regulatory landscape. For each category, a combination of technical and organizational controls are in place, with responsibilities shared among Tezos core developers, validators, custodians, and institutional product operators.

\subsection{Key Risk Scenarios and Mitigations}
\paragraph{R1. Protocol Implementation Flaws.}\label{risk:R1} Because the system is enshrined in the Tezos protocol, a flaw in its code could potentially impact network consensus or the solvency of the aggregated stake.
\begin{itemize}
	\item \textbf{Mitigation:} The code is subjected to multiple independent security audits prior to activation, and changes are exercised on multiple public \glspl{testnet} during each phase of the protocol governance process. Core engineers provide continuous monitoring and maintenance. New features must go through the full on-chain amendment process, while targeted bug fixes can be shipped on a shortened path consistent with the \href{https://octez.tezos.com/docs/developer/protocol_playbook.html}{Tezos Protocol Playbook}.
\end{itemize}

\paragraph{R2. Malicious Validator Behavior.}\label{risk:R2} An attacker could register new "throwaway" validator entities, offer zero-fee services to attract a large allocation from the system, and then intentionally cause a slashable offense (like double-baking) to harm the system.
\begin{itemize}
	\item \textbf{Mitigation:} The \emph{\ProjectName{} protocol} enforces several rules to counter this: stake allocations to any single validator are capped relative to their own security deposit (self-bond), validators must have a clean recent history before becoming eligible, and any validator that is slashed is immediately and automatically excluded from future allocations. Slashing on Tezos is extremely rare; since the Paris activation and adoption of Adaptive Slashing, only a single slashing event has been recorded. Under Adaptive Slashing, penalties remain very small unless many committee members collude or a single validator controls a very large share of stake. Losses, if any, are socialised across the system via a small step-down in $R_t$.
\end{itemize}

\paragraph{R3. Validator Concentration.}\label{risk:R3} A large, dominant validator could try to capture a significant share of the system's stake by offering unsustainably low fees.
\begin{itemize}
	\item \textbf{Mitigation:} The \ProjectName{} protocol's stake allocation algorithm enforces dynamic, per-validator caps to prevent any single entity from accumulating an excessive share. Eligibility rules (clean history, capacity, declared fees) also reduce capture risk in practice.
\end{itemize}

\paragraph{R4. Market Liquidity Stress.}\label{risk:R4} A sudden surge in redemption demand or a major dislocation in the secondary market could cause the price of \TokenTicker{} to temporarily trade at a discount to its fair value.
\begin{itemize}
	\item \textbf{Mitigation:} The ability for arbitrageurs to always mint and redeem at the intrinsic value provides a strong anchor for the market price. Additionally, operators of regulated products are encouraged to maintain their own supplemental liquidity buffers to handle extraordinary withdrawal requests, consistent with the system's design.
\end{itemize}

\paragraph{R5. Governance Attacks.}\label{risk:R5} While the amendment process is robust, a coordinated group of stakeholders could theoretically propose and pass a harmful change to the protocol.
\begin{itemize}
	\item \textbf{Mitigation:} Passing an amendment requires \gls{supermajority} across multiple voting phases with weeks of notice, making malicious changes extremely unlikely. If a harmful change were somehow activated and identified, a corrective protocol override (bug-fix amendment) would be prepared and activated on an expedited basis.
\end{itemize}

\paragraph{R6. Data Unavailability or Manipulation.}\label{risk:R6} NAV agents and other third parties depend on accurate, timely data from the blockchain.
\begin{itemize}
	\item \textbf{Mitigation:} All core data is stored publicly on-chain and broadcast through events, allowing for redundant, independent indexers. Product issuers are expected to reconcile the core data points ($\LedgerBalance{}$ and $\TokenSupply{}$) from multiple independent sources before publishing official NAV figures.
\end{itemize}


\subsection{Security and Resilience Program}
Several ongoing workstreams support protocol security and day-to-day resilience:
\begin{itemize}
	\item \textbf{Engineering reviews:} Independent audits, peer review, and post-mortems drive fixes and improvements. The codebase uses OCaml’s type safety, property-based testing, and model-based testing to surface regressions early.
	\item \textbf{Runbooks for incidents:} Documented procedures cover slashing events, large redemption queues, and communication steps for institutional partners.
	\item \textbf{Monitoring:} Automated alerts flag unusual allocation changes, queue spikes, or abrupt movements in $R_t$.
	\item \textbf{Validator hygiene:} Operators are encouraged to publish contacts and infrastructure attestations; eligibility remains permissionless and rules-based.
\end{itemize}

\subsection{Custody and Key Management}
The protocol is non-custodial and has no privileged administrator keys. For institutional users, managing \TokenTicker{} tokens is no different from managing standard Tezos assets. Custodians can use their existing, SOC-certified workflows for key management. Advanced security measures like segregated accounts, address allowlisting, and \gls{mpc} can be layered on top without requiring any changes to the underlying protocol.

\subsection{Slashing and Loss Handling}
In the event of a slashing, the associated loss is immediately and transparently reflected in the \TokenTicker{} exchange rate. This removes any ambiguity about liability. The protocol's on-chain state clearly identifies the faulty validator, the cycle in which the event occurred, and the amount of the loss, providing a clear audit trail for investor statements and regulatory disclosures. Any denunciation rewards recovered from the slashing event are automatically credited back to the system, partially offsetting the loss.

\subsection{Operational Resilience and Disaster Recovery}
The entire state of the system can be fully reconstructed using only the data on the Tezos blockchain. Any actor can run an archive node and bootstrap from \gls{genesis} to independently reconstruct all events and current state. Reference indexers and monitoring scripts will be open-sourced, with documentation to guide custodians and NAV agents through disaster recovery scenarios. Coordination with the Tezos Foundation and other major ecosystem participants ensures multiple, redundant sources of data are always available.

\subsection{Market Dynamics (Outstanding)}

\medskip
The market behaviour of \TokenTicker{} is anchored by its intrinsic, protocol-defined exchange rate $\ExchangeRate{}$, yet secondary market prices can deviate temporarily due to liquidity, flows, and frictions:
\begin{itemize}
	\item \textbf{Price anchor and arbitrage.} Primary mint/redeem at intrinsic value creates a strong arbitrage band. Discounts tend to close as arbitrageurs buy in the market and redeem, while premiums are capped by the ability to mint and sell.
	\item \textbf{Flow-driven dislocations.} Large, one-sided flows (e.g., redemptions) can push the token to a temporary discount when market depth is thin. As queues are cleared and arbitrage acts, the price converges back toward intrinsic value.
	\item \textbf{Yield sensitivity.} The carry of holding \TokenTicker{} increases with the underlying staking yield on XTZ. Higher expected yield generally supports tighter discounts (or transient premiums) in risk-on regimes; the anchor remains the intrinsic rate.
	\item \textbf{Event handling.} In a slashing event, the intrinsic rate $R_t$ steps down immediately and transparently. Efficient markets should reflect that step change swiftly; clear communications and automated disclosures reduce uncertainty premia.
	\item \textbf{Liquidity provisioning.} Deeper on-chain liquidity and designated market makers on centralised venues both reduce volatility around the anchor. Protocol transparency (on-chain $\LedgerBalance{}$ and $\TokenSupply{}$) lowers informational frictions for NAV agents.
	\item \textbf{Competitive landscape.} Multiple liquid staking options can fragment liquidity. Being protocol-enshrined reduces smart-contract risk, which can improve institutional adoption and, over time, concentrate liquidity in \TokenTicker{}.
\end{itemize}

\section{Governance}
\label{sec:governance}

\subsection{Amendment Lifecycle (Summary)}
As a core component of the Tezos protocol, \ProjectName{} is governed by the established amendment process (see \href{https://docs.tezos.com/architecture/governance}{Tezos governance architecture}) and the original Tezos whitepaper~\cite{tezos_whitepaper}. This formal, multi-stage procedure ensures that upgrades are deliberate, transparent, and well-communicated. Proposed changes pass through five fixed periods (14 cycles each, roughly 2.5 months in total); if any vote fails quorum or \gls{supermajority}, the process restarts at the proposal period.
\begin{itemize}
  \item \textbf{Proposal period.} Validators submit and upvote protocol \glspl{hash} (each hash identifies a tarball of protocol source files). Approval voting picks the hash with the most support, provided participation reaches the proposal quorum; otherwise the cycle restarts.
  \item \textbf{Exploration vote.} Validators cast a single \textit{Yea/Nay/Pass} ballot weighted by stake. Meeting quorum participation and an 80\% super-majority of Yea moves the candidate forward; failure resets to proposal.
  \item \textbf{Cooldown period.} No vote takes place. Teams continue reviewing, testing, and shipping documentation or tooling ahead of activation.
  \item \textbf{Promotion vote.} Validators run the same \textit{Yea/Nay/Pass} ballot. Quorum plus 80\% super-majority approves; otherwise the amendment process reverts to the proposal period. (Tezos updates the quorum (minimum participation) via an \gls{ema} (exponential moving average) of past participation.)
  \item \textbf{Adoption period.} Operators finish upgrades. The protocol activates automatically at the end of the period and a new proposal window opens.
\end{itemize}

\subsection{Immutable Core Parameters}
The fundamental rules of the system are expressed as protocol constants that can only be changed through the formal amendment process described above. These parameters include:
\begin{itemize}
    \item Per-validator stake allocation caps.
    \item Global limits on the total aggregated stake size.
    \item Eligibility criteria for validators to receive stake allocations.
    \item The configuration of the competitive fee market.
    \item Policies governing the redemption of \TokenTicker{} tokens.
\end{itemize}
Crucially, there is no "emergency administrator" or multisig group with the power to bypass this process. This design choice reinforces the system's neutrality and guarantees that its core mechanics cannot be altered arbitrarily, while still allowing for deliberate, community-approved evolution over time.

\subsection{Clear Communication for Stakeholders}
To support institutional users, all official documentation related to a protocol proposal—such as descriptions on Tezos Agora or on-chain metadata—will explicitly detail any potential impacts on \ProjectName{}. Furthermore, governance-related communications will include annexes specifically for institutional stakeholders, summarizing:
\begin{itemize}
    \item The expected timeline for the proposed changes.
    \item Any required operational adjustments for custodians or NAV agents.
    \item Notice templates that regulated products can adapt for their own investor communications.
\end{itemize}

\subsection{Transparency and Accountability}
The entire governance process is public. Vote counts, validator participation rates, and the rationales behind voting decisions are all available for public review. To further enhance accountability, governance dashboards will be maintained to highlight the voting records of validators who receive allocations from the \ProjectName{} system. This allows institutional participants to assess whether the validators they rely on are acting as responsible stewards of the protocol. After any upgrade is activated, post-mortems will be published to analyze the observed effects, with the findings informing future governance proposals.

\subsection{Voting Power and Future Extensions}
\ProjectName{} assigns stake to validators for block production and attestation, but it does not add to their governance voting power: amendment ballots remain bound to each validator's own bonded stake, plus any tez staked or delegated to them directly outside \ProjectName. This keeps \ProjectName{} from concentrating amendment influence while still providing operational scale for validation.

Looking ahead, we intend to explore contract-level voting hooks so \TokenTicker{} holders can cast protocol ballots directly on their staked tez. Enabling token holders to vote without routing through validators would make governance more open and equitable, while preserving the simplicity of aggregated staking.
\section{Roadmap}
\noindent\textit{Dates are indicative and assume successful governance approval.}
\begin{itemize}
	\item \textbf{Q4 2025:} Project kickoff; align protocol, product, and legal workstreams; initiate enshrined contract specification.
	\item \textbf{Late Q1 2025:} Finalise prototype ledger and \TokenTicker{} ledger; complete UX and indexer integration tests; commission external security reviews.
	\item \textbf{February 2026:} Publish white paper v1; enable feature-flagged \gls{testnet} release; deliver standardized reporting specifications 
	      (NAV procedures, disclosure templates).
	\item \textbf{March--April 2026:} Run economic simulations and red-team staking/flash-unstake scenarios; refine allocation parameters; execute community and institutional feedback sessions.
	\item \textbf{Q2 2026:} Public testnet evaluation of redemption queue performance and per-cycle bucket stress tests; publish benchmarking report.
	\item \textbf{September 2026:} Target mainnet activation, subject to governance approval and completion of all outstanding deliverables.
\end{itemize}

\noindent\textit{Milestones and dates are indicative and non-binding; actual sequencing depends on governance and external reviews.}

\section{Glossary}
\label{sec:glossary}
\glsaddall[types={protocol}]
\glsaddall[types={mechanics}]
\glsaddall[types={finance}]
\glsaddall[types={regulatory}]
\glsaddall[types={tech}]

\printglossary[type=protocol,nonumberlist]
\printglossary[type=mechanics,nonumberlist]
\printglossary[type=finance,nonumberlist]
\printglossary[type=regulatory,nonumberlist]
\printglossary[type=tech,nonumberlist]


\bibliographystyle{plainnat}
\bibliography{references}

@misc{tezos_whitepaper,
  author = {L .M Goodman},
  title = {Tezos: a Self-Amending Crypto-Ledger},
  year = {2014},
  url = {https://tezos.com/whitepaper.pdf}
}
\appendix
\section*{Appendices}
\addcontentsline{toc}{section}{Appendices}
\section*{A. Core Operations}
\begin{itemize}
  \item \apicode{stake(amount)}: A user deposits tez into the system and mints a corresponding amount of \TokenTicker{}.
  \item \apicode{request\_unstake(amount)}: A user burns \TokenTicker{} units to initiate the unstaking process, creating a claim on frozen tez.
  \item \apicode{finalize\_unstake(ticket)}: After the unbonding period, the user can claim their finalized tez using the redemption ticket.
  \item \apicode{register\_validator(fee, capacity)}: A validator opts-in to the system, specifying their fee and the maximum stake they can take.
  \item \apicode{update\_validator\_parameters(fee, capacity)}: A validator can update their fee and capacity.
  \item \apicode{unregister\_validator()}: A validator can opt-out of the system.
\end{itemize}

\section*{B. RPC Interface}
The protocol will expose a set of RPC endpoints to query the state of the liquid staking system, including:
\begin{itemize}
    \item \apicode{/ledger/state}: Returns the total amount of tez in the system, the total supply of \TokenTicker{}, and other global parameters.
    \item \apicode{/ledger/validators}: Returns a list of all registered validators with their fees, capacities, and current stake.
    \item \apicode{/ledger/allocations?cycle=\{n\}}: Returns the stake allocation for a given cycle.
    \item \apicode{/user/\{address\}/balance}: Returns the \TokenTicker{} balance for a given user.
\end{itemize}

\section*{C. Testing \& Validation}
The implementation will be validated through a series of tests, including:
\begin{itemize}
    \item Unit tests for all individual functions.
    \item Integration tests simulating the complete lifecycle of staking, unstaking, and reward distribution.
    \item Property-based tests to verify the system's invariants under a wide range of inputs.
    \item Simulation campaigns on a public \gls{testnet} to validate the economic incentives and behavior of the system under real-world conditions.
\end{itemize}


\section*{D. Reference Model (Python)}
\label{app:nav-reference}
\paragraph{Illustrative Python translation}
  extit{Note: This Python is illustrative (\gls{offchain}) and models the same logic as the pseudocode above. Naming uses “allocation/assigned” (not “delegation”) in line with house style.}

\vspace{0.25em}
\noindent\textbf{Stake Allocation (Python)}
\begin{lstlisting}[language=Python,firstline=12,lastline=49,style=AppendixCode]
@dataclass(frozen=True)
class ValidatorProfile:
    """Simplified validator profile used in the greedy allocation example."""
    address: str
    fee: float  # e.g., 0.05 for 5%
    capacity: int
    current_stake: int
    eligible: bool = True

def allocate_stake(
    validators: Iterable[ValidatorProfile],
    total_stake: int,
    *,
    per_validator_cap_fraction: Optional[float] = None,
) -> List[Tuple[str, int]]:
    """Allocate stake to eligible validators greedily (lowest fee first)."""
    
    eligible = [v for v in validators if v.eligible]
    sorted_validators = sorted(eligible, key=lambda v: (v.fee, v.address))
    
    remaining = total_stake
    allocations: List[Tuple[str, int]] = []
    
    for validator in sorted_validators:
        if remaining <= 0:
            break
            
        fraction_cap = int(per_validator_cap_fraction * total_stake) if per_validator_cap_fraction else validator.capacity
        limit = min(validator.capacity, fraction_cap)
        
        headroom = max(0, limit - validator.current_stake)
        allocation = min(headroom, remaining)
        
        if allocation > 0:
            allocations.append((validator.address, allocation))
            remaining -= allocation
            
    return allocations
\end{lstlisting}

\noindent\textbf{Redemption Logic (Python)}
\begin{lstlisting}[language=Python,firstline=51,lastline=84,style=AppendixCode]
@dataclass
class LedgerState:
    """Mutable protocol state used by exchange-rate helpers."""
    balance: DecimalLike  # L_t (Active Ledger Balance)
    supply: DecimalLike   # S_t (Token Supply)
    frozen: DecimalLike = 0 # F_t (Frozen/Unbonding Ledger)

    def rate(self) -> Decimal:
        return self.balance / self.supply if self.supply > 0 else Decimal(1)

def request_redemption_immediate(state: LedgerState, units_to_burn: DecimalLike) -> Decimal:
    """
    Implements 'Immediate Burn' (Section 2.5).
    Burns tokens immediately, calculates tez value, and moves it to Frozen Ledger.
    """
    units = _decimal(units_to_burn)
    if units <= 0:
        return _decimal(0)

    # 1. Calculate value using CURRENT rate
    # v = u * R_t
    tez_value = units * state.rate()

    # 2. Atomic State Update
    # S_t := S_t - u
    state.supply -= units
    
    # L_t := L_t - v
    state.balance -= tez_value
    
    # F_t := F_t + v (Move to frozen ledger)
    state.frozen += tez_value

    return tez_value
\end{lstlisting}

\end{document}